# AKN_Regie: Bridging Digital and Performing Arts

Georges Gagneré
INREV-AIAC
Paris 8 University
Saint-Denis - France
georges.gagnere [at] univ-paris8.fr

Anastasiia Ternova
INREV-AIAC
Paris 8 University
Saint-Denis - France
anastasiia.ternova [at] gmail.com

## ABSTRACT

AvatarStaging framework consists in directing avatars on a mixed theatrical stage, enabling a co-presence between the materiality of the physical actor and the virtuality of avatars controlled in real time by motion capture or specific animation players. It led to the implementation of the AKN_Regie authoring tool, programmed with the Blueprint visual language as a plugin for the Unreal Engine (UE) video game engine. The paper describes AKN_Regie main functionalities as a tool for non-programmer theatrical people. It gives insights of its implementation in the Blueprint visual language specific to UE. It details how the tool evolved along with its use in around ten theater productions. A circulation process between a non-programming point of view on AKN_Regie called Plugin Perspective and a programming acculturation to its development called Blueprint Perspective is discussed. Finally, a C++ Perspective is suggested to enhance the cultural appropriation of technological issues, bridging the gap between performing arts deeply involved in human materiality and avatars inviting to discover new worlds.

## 1 Introduction

At a time when the digital revolution seems to have touched every sector of society and every country on the planet, its impact on our lives and socio-cultural practices has probably not yet produced its full effects. The paradigm shift induced by the emergence of computing and its electronic deployment linked to the mastery of electricity probably exceeds the measure of the previous paradigm shift that characterized the emergence of the alphabet, writing and the techniques of writing around the first millennium BC [29]. Fully immersed in the culture of literacy, it is hard to imagine human beings with equivalent cognitive capacities but without recourse to the written word to communicate and develop. This culture of literacy, which took several centuries to emerge and requires several years to acquire, is currently confronted with new practices based on a new way of representing reality and the acquisition of new communication skills using computer code. [23] hypothesized that the performances of ancient Greek tragedies, among the earliest written traces of the theatrical art form, produced by Aeschylus, Sophocles or Euripides, had promoted the acculturation of Greek society to the alphabet and writing. This would have given rise to an individual consciousness characteristic of this civilization. "The Greek stage projected the prototypes of Western man as models for the acquisition of private consciousness." The influence of technology on the cultural evolution of societies has been studied by philosopher and epistemologist Gilbert Simondon. "Culture is that by which man regulates his relation to the world and his relation to himself; yet, if culture did not incorporate technology, it would contain an obscure zone and could not bring its regulating normativity to the coupling of man and the world." [36]

In Simondon's wake, this paper proposes to study the cultural appropriation of computing and communication technologies in the context of a dialogue between digital arts and performing arts practices. The figure of the actor delineates boundaries between the materiality of the physical actor and the virtuality of the avatar. We introduce AKN_Regie, a software authoring tool dedicated to directing avatars on a mixed reality theatrical stage and accessible to non-programmer theater technical managers and artists.

Section 2 introduces related works. Section 3 describes AKN_Regie main functionalities as a plugin of Unreal Engine (UE) video game environment developed by Epic Games [11]. Section 4 gives insights of its implementation in the Blueprint visual language specific to UE. Section 5 details how the tool evolved along with its use in around ten theater productions. Section 6 discusses the circulation process between a non-programming point of view on AKN_Regie called Plugin Perspective and a programming acculturation to its development called Blueprint Perspective. We finally suggest that a C++ Perspective could enhance the cultural appropriation of technological issues, bridging the gap between performing arts deeply involved in human materiality and avatars inviting to discover new worlds.

## 2 Related works

In 1965, with the choreographic collaboration of Merce Cunningham and the technological input of Billy Klüver, engineer at BELL Laboratories, John Cage created *Variation V*, in the frame of a performance cycle initiated in 1958 [27]. This is one of the first examples of interaction between dancers and a musical composition, which they were modifying through their movements captured by magnetic antennas designed by the BELL Laboratories engineer. The music followed the dancers, rather than the opposite, opening new creative territories to be explored using emerging technologies. In 1966, Cage and Klüver repeated the experiment, involving nine other artists, including Robert Rauschenberg and Robert Whitmann, and several dozen engineers, also from BELL Laboratories, including Max Mathews and Fred Waldhauer. Several months of work resulted in *9 evenings: Theater and Engineering* [30], the founding event in the collaboration between art, technology, and science, consisting of the public presentations during 9 evenings of the artists' performances, including John Cage's *Variation VII*. In 1967, engineers Klüver and Waldhauer and artists Rauschenberg and Whitmann founded an association called *Experiments in Art and Technology* [26], officially formalizing the desire for collaboration shared by artists and engineers.

[35] has drawn up an overview of collaborations between the arts and new technologies, highlighting the creative entanglement between artistic and technological dimensions. Recently, the *Nautilus* show [12] followed in the footsteps of Klüver, who retrospectively summed up his experiences of the 60s by explaining that the "goal from the beginning was to provide new materials for artists in the form of technology" [2], enabling the engineer to take advantage of stimulating experimental situations to solve technological problems. Amongst numerous examples we could also quote the collaboration between the French company Dassault System and a prima ballerina, which gave rise to the show *M. & Mme Rêve* in 2013 as a tribute to Ionesco [8], or the collaboration between the famous Shakespeare Company and the global microprocessor manufacturer DELL, which together produced Shakespeare's *Tempest* in 2016.

Genuine interdisciplinary collaboration between artists, technological experts and scientists often poses methodological problems. In the United States, Nicholas Negroponte founded the Media Lab in 1985 at the Massachusetts Institute of Technology, a research center dedicated to technology and design, in response to the emerging challenges of Art-Technology collaborations. Among the many artistic software programs created by Media Lab students may be mentioned Processing, developed by Benjamin Fry and Casey Reas [34]. In Europe, Franck Popper, one of the inventors of the Plastic Arts concept in French higher education, co-founded the corresponding department at the Université expérimentale de Vincennes in the late 60s, which later became Paris 8 Vincennes-Saint-Denis University. As early as 1975, he devoted particular attention to the relationship between art, science, new technologies and creativity in his seminal work *Art, Action and Participation* [32]. In 1984, he supported the creation of the Arts et Technologies de l'Image (ATI) faculty in the visual arts department of Paris 8 university, associated with the Images Numériques et Réalité Virtuelle (INREV – Digital Images and Virtual Reality) research team, which is a pioneer in training art students in both art and technology, particularly computer science. Co-founder of ATI, Edmond Couchot went on to theorize the notion of digital art in France [6], and throughout his career has underlined the need for collaboration between the digital and living arts, notably through interdisciplinary co-creations involving artists and technological experts [5].

In the literature documenting art-science-technology projects, the authors of [22] highlight the need for a common vocabulary and propose as a solution to share a common research object from the very beginning of the project, without artistic or technological prejudice. "We regularly used to express the need of a common vocabulary to understand the multiple and seemingly divergent emerging points of view. But the success in collaboration came from fully sharing a common object and its setups, a digital shadow interacting with a virtual world in an augmented environment".

Also noting a "need for a common ground for dialogue", the authors of [3] propose "to use a "proxy", a person who has both artistic and technical abilities." [33] points out that a collaboration between art and technology "is more than the sum of the individual parts; interdisciplinarity does not consist of doing isolated work and then putting some glue between the results, but really finds a way of working together, mixing methodologies, expectations, dissemination channels, etc. of different disciplines." In turn, she takes up the figure of the proxy, "a person who has both artistic and technical abilities in order to overcome the difficulties of collaborating". In her description of a creation research project, [31] analyzes a collaboration between a director and a programmer and "speak of gradual "transitions from one discipline to another" to build bridges between various domains of knowledge and practice. As a result, the programmer becomes more than just an IT specialist, subservient to the demands of the director, who in turn is more than just an artist, presenting his concept in the obscure language of art. "Boundary crossing" is an interesting methodological proposition which will most probably bring good results in other interdisciplinary projects."

Myron S. Krueger prefigured the notion of a "proxy" with both computer and artistic skills, able to manipulate computer tools to create devices for artists and audiences. In 1977, describing his *Videoplace* work, which gives a central place to the spectator interacting with computer-generated images, he declared that "the responsive environment has been presented as the basis for a new aesthetic medium based on real-time interaction between men and machines." [24] He clarified in 1998 that his project had been from the outset "to raise interactivity to the level of an art form as opposed to making artwork that happened to be interactive." [25] So it is not a question of putting technology at the service of art, but of proposing artistic devices based on the expressive potential of technology [35, p. 320]. We should also remember the figure of Max Mathews, who put himself at the service of artists on *9 evenings*, but who also created software since 1957 to enable musical composition. It was in honor of his contribution that the Max/MSP software [7], a widely used tool in the musician-programmer community, was named after its creation by Miller Puckett at the Institut de Recherche et Coordination Acoustique/Musique (IRCAM) in Paris in the 1980s.

The authors of [37] elaborate recommendations based on lessons learned from several projects involving collaboration between art and culture professionals and programmers. The need for a common ground for dialogue remains a central issue. Programmers "should not hesitate to bring in their own ideas. Artists are recommended to work at a deep understanding of the technologies involved." In practice, this means starting an introduction to software requiring programming skills, such as video game software. Another way of bridging the gap is to create digital tools that artists can program themselves, and which can be improved from project to project. "The developers did not implement the artists' ideas directly, but provided the tailored tools, with the help of which the artists created prototypes of the technical artwork on their own."

Whether in the creative appropriation of complex Augmented Reality/Virtual Reality (AR/VR) tools such as video game engines, or in their use to create authoring tools, the authors of [28] note three main difficulties for non-programmer practitioners. After identifying several classes of tools, they note "(1) a massive tool landscape […] (2) most AR/VR projects require tools from multiple classes […] (3) significant gaps of tools both within and between classes of tools." They point the possibility of integrating accessible prototyping tools into complex software. "More complex AR/VR interfaces, however, can involve lots of interactive 3D objects reacting to both users' explicit interactions (via touch, gesture, speech) and implicit interactions (via device camera, inertial sensors, external sensors)." The authors of [1] confirm the growth of users wishing to develop authoring tools without formal learning in computer science, whether hobbyists or domain experts. After identifying three types of barriers in understanding the AR/VR landscape, in designing and prototyping AR/VR experiences and in implementing and testing AR/VR applications, they suggest that these people "can find a better match between expressivity and learnability – end user programmers in AR/VR can benefit from starting with a simple development environment but with the opportunity to learn the more advanced concepts directly inside the tool. One way to do this could be to draw upon the adaptive interfaces literature to tailor feature-rich interfaces of complex authoring environments according to users' expertise level."

## 3 Technological context and tool functionalities

### 3.1 AvatarStaging

AvatarStaging [17] is a framework which combines several software tools to transpose movements performed by an actor equipped with a motion-capture device – the "mocaptor" – onto the avatars inside a virtual theater, which represents the virtual component of the mixed-reality stage.

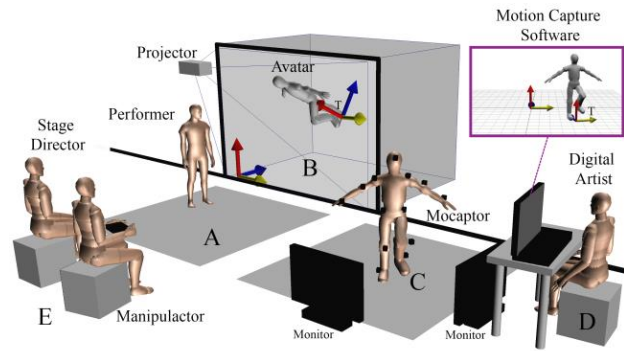

**Figure 1: AvatarStaging setup**

AvatarStaging brings together three main operations, that let a mocaptor control an avatar and develop a relationship between said avatar and a physical actor on a mixed-reality stage. Figure 1 is a schematic representation of how these three operations work for a mocaptor, an avatar, and an actor, variations of which can be inferred in more complex configurations. The first operation is capturing the motion itself. The mocaptor wears a full-body suit, which can be optical or inertial, and which makes it possible to collect data relating to the position of the sensors situated on specific areas of the mocaptor's body (Figure 1, zone C). The motion-capture data are transferred to a dedicated software where a virtual skeleton is inferred from the position of the sensors and streamed to other software. The second operation, called motion retargeting, involves transposing the movements of the first skeleton onto the skeleton of the avatar chosen to perform in the 3D space in dialogue with the actor. The avatar has usually a specific skeleton distorting the mocaptor's original motions. The third and last operation consists in organizing the positions of the avatar and/or its prerecorded animations according to the digital setting and to the positions of the physical actor (Fig. 1, B), using AKN_Regie.

Consequently, the mocaptor present in space C develops relationship to the actor on stage A thanks to the video feedback monitors placed around him, which enable him to see the result of his movements projected onto his avatar in digital space B. This dynamic is made possible by the three operations of motion capture, motion retargeting, and motion control and involves, on the software side, the following requirements, brought together in the AKN_Regie tool:

- Retrieve data from real-time motion capture devices and apply it to avatars
- Precisely place avatars in 3D space and orient them with game controllers
- Program a sequence of successive configurations for manipulating avatars in 3D space
- Use the keyboard or controllers to manipulate the pipeline during a performance

The real-time dimension of motion capture and interaction with a physical actor led to developments being made in the UE video game engine, mainly chosen because of its Blueprint visual programming functionality, which allows developing without having to write code in script form.

## 3.2 Using AKN_Regie by manipulating blueprints

UE is developed in C++, and the source code is accessible, modifiable and compilable. Programming a game with the engine can be done in two ways: using the C++ language in script form, or using a kind of transposition of the C++ language in visual form, specifically developed in UE by Epic Games under the name of Blueprint [10]. By extension, the name blueprint also designates each piece of code programmed with the visual Blueprint language, made up of programming blocks called nodes and arranged in a graph, called EventGraph.

Figure 2 shows the blueprint interface editor, with the example of the Regie_Cuesheet blueprint created with the AKN_Regie tool. A non-IT user can manipulate the blue boxes (corresponding to the nodes) (figure 2B) and set up a pipe in the form of a menu in the right-hand column (figure 2A).

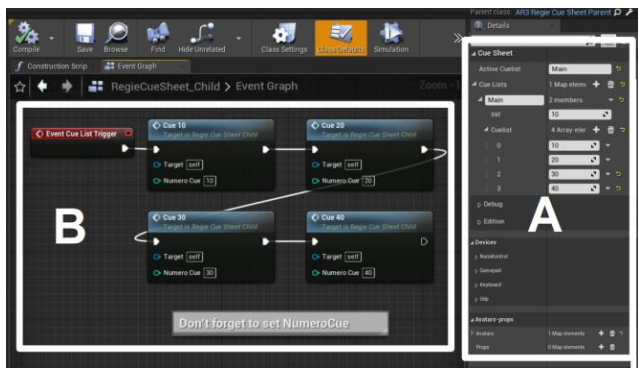

**Figure 2: Regie_Cuesheet blueprint. B (left): cue edition, A (right): control parameters**

UE follows the programming principles of video game engines, which are based on a graphical interface editor that lets you access 3D space, called a level. In the construction of a video game, there are various ways of generating necessary elements for active navigation in the level: they can be placed directly in the level using the main interface editor, but they can also be programmed to be spawned at any time during the game, and notably at launch, by indicating where they should be spawned. In most video games, the 3D objects that make up the fixed scenery are placed directly in the level. The player character, on the other hand, is spawned on its starting position (player start).

We proceed in the same way with AKN_Regie, building the scenery directly in the level, and giving the user the option of casting characters from a list of predefined avatars prepared to be manipulated by the main AR3_Regie blueprint. This cast of avatars is then spawned at game launch. We also ask the user to place in the level GoalAvatar objects (GA) at the successive locations where he will position the avatars (figure 3). The set of GAs in a level corresponds to the blocking of the avatars cast. The same operations apply to the placement of props he wishes to manipulate in the scenery, with the positioning of GoalProp objects (GP), and to the positioning of the camera that will render the desired views on the level, with GoalCamera objects (GC).

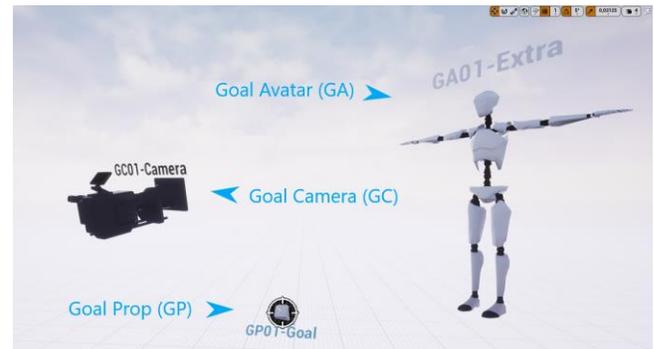

**Figure 3: The 3 goal objects in a level: camera, avatar and prop**

The user must therefore build a scenery in a level, drag and drop into it the Regie_Cuesheet blueprint and the three types of goals for all the desired positions of the avatars, props and camera. The goals are therefore essentially markers with the appearance of the objects whose positions they indicate in the level. There are two types of props: Autonomous props can be placed directly in the level. They can also be attached to a part of an avatar's body as Dependent Props. Their nature is limited only by the imagination of the designers and the capabilities of the programmers. The current version of AKN_Regie offers as props a mesh, a light source, particles, or an audio file.

## 3.3 Organizing the Regie_Cuesheet

The second step is to organize the Regie_Cuesheet. This key blueprint is like a virtual stage manager, synchronizing all the dynamic elements of the staging at each cue, corresponding, for example, to the appearance or change of appearance of a prop, the triggering of an avatar animation, and more generally to any modification of objects that can be manipulated with AKN_Regie.

The Regie_Cuesheet consists of two parts. The first, Part A, is dedicated to defining the avatar and props objects that will be active in the staging, linking these virtual objects to the physical manipulation devices available (keyboard, joystick, MIDI controller), adjusting operating parameters (gamepad speed for instance) or defining the cue order (figure 2A). Figures 4 and 5 focus on three sections of frame A in figure 2.

Figure 4 on the left shows that this example uses 3 cues numbered 10, 20 and 30 in the Main cuelist as well as a Cinecamera. In a Regie_Cuesheet, several cuelists can be created to launch cues in a different order. Figure 4 on the right corresponds to the Devices section and shows the use of a NanoKontrol midi controller and two gamepads. This is where indicating to which avatar or prop each of the gamepads will be assigned. Figure 5 illustrates the Avatars-props section and shows the use of two different types of avatars (Avatar1 and Avatar2) and two different autonomous props (Prop1 and Prop2). We have also chosen to assign a dependent prop (Prop3) to Avatar1's left arm.

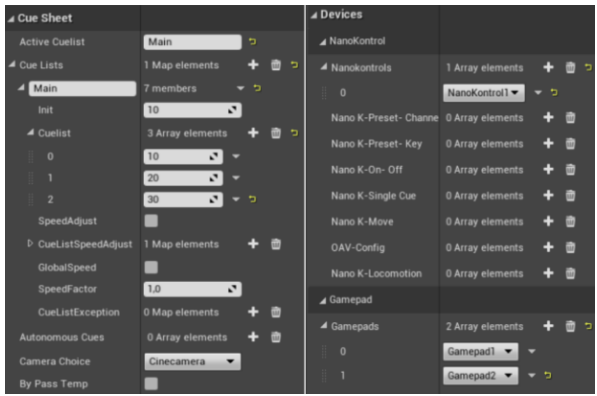

**Figure 4: CueSheet panel (left) and Devices panel (right)**

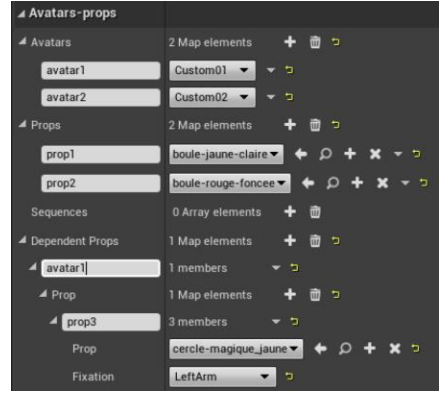

**Figure 5: Avatars-props menu**

The second part B is a graphical area giving access to all the cues in a show (figure 2B). It provides access to each cue so that they can be configured when each box is clicked. This brings to a new window (figure 6) to organize desired actions using parameter tables called "Sets". The use of Sets in AKN_Regie is quite intuitive. All Sets include a Bypass feature to ignore the changes indicated in the Set. This feature is useful for projects that rely on experimentation and improvisation and need activating or deactivating Sets on the fly.

In the Camera, Prop and Avatar Sets, there is a goal parameter, which corresponds to the element's position in 3D space. The ability to quickly define and change goals allows to adjust interactions precisely and quickly between virtual elements, as well as between physical actors and virtual elements, including their mutual positions to ensure a sense of presence. Fast jumps between goals also offer the possibility of easily creating all kinds of comic, magical or dramatic effects. Finally, this function offers ease of use in the rehearsal or improvisation process. The other Sets functions are adapted to the type of each element: for instance, creating a fade in or fade out for the camera, adjusting the appearance of the avatar and the type of animation, or specifying the prop parameters.

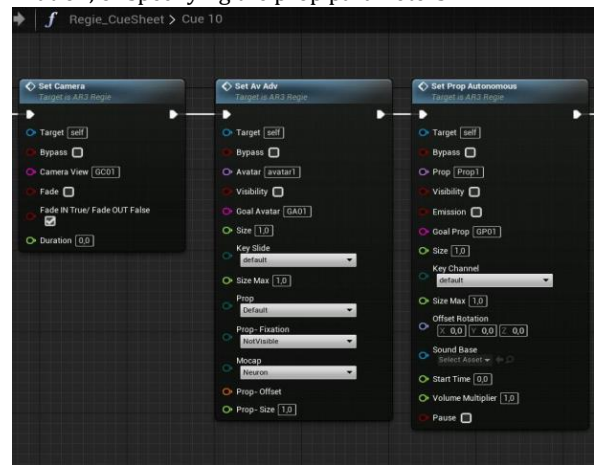

**Figure 6: Sets parameters in Cue10**

Placing goals in a level in a first step and organizing the Regie_Cuesheet blueprint in a second step are accessible to non-programmer technical manager and artists as section 5 shows it through short introductions of performances based on AvatarStaging framework and its AKN_Regie authoring tool. We propose now to contextualize the tool in its programming environment by giving some insights on its architecture.

## 4 AKN_Regie programming architecture

Usually, during a video game programming, a level is launched in game mode for testing by the programmer. And when all the programming has been completed and validated, the game is packaged in the form of a standalone, which is launched directly in game mode by the player and no longer requires the global programming environment. A standalone represents a set of files weighing a few hundred MB, whereas the video engine programming environment represents tens of GB of files.

The current version of AKN_Regie requires the UE interface editor and programming environment, which must be installed beforehand. After setting the tool to manipulate avatars as described in the previous section, the user launches the engine in game mode and visualizes the results in the desired level. There is currently no AKN_Regie standalone, but the installation of the UE programming environment is accessible to non-programmer user.

### 4.1 Blueprint development in a plugin

To facilitate accessibility to the many programming possibilities offered by UE, development of AKN_Regie is also carried out in the blueprint language, limiting itself to the engine nodes that are part of the central kernel present by default at installation (Built-In Engine). It requires only the activation of an additional plugin for MIDI event management. Acronym for the audio communication protocol Musical Instrument Digital Interface, MIDI is a widespread protocole to connect controllers to software interfaces.

The notion of plugins in UE programming environment consists in autonomously compartmentalizing code, which can then be easily added to an existing project and give access to new nodes to perform specific actions. For example, a motion capture device plugin can be added, which will then receive data from the matching device connected to the computer. AKN_Regie has therefore been placed in a plugin that can be added to any existing project to use the tool with the desired avatars. Thanks to this feature, the tool can be used on different projects.

The AKN_Regie plugin contains a collection of blueprints. The main one, AR3_Regie, shares the same programming environment as when a non-programmer uses AKN_Regie with a simplified blueprint of the type shown in figure 2. All the nodes used are updated by default with each new version of the engine, thus ensuring the consistency of the plugin as the engine evolves.

### 4.2 C++ inheritance

Without going into too much details about the characteristics of C++ programming, we can briefly introduce the fundamental notion of the class, which makes it possible to compartmentalize programming elements and spawn so-called children, which benefit from the functionalities of the parent class, allowing reorganization and implementation of new specifications. Code can be "hidden" to simplify its use. This parenting principle links AR3_Regie parent blueprint (figure 7) to Regie_Cuesheet child blueprint (figure 2).

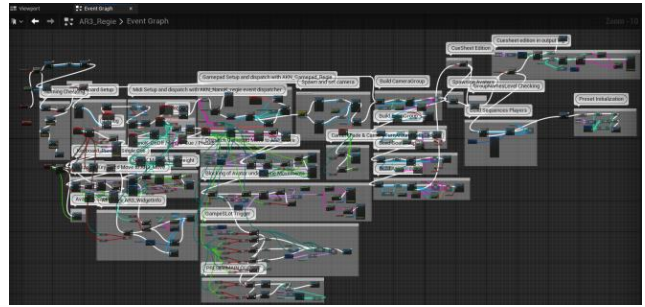

**Figure 7: AR3_Regie blueprint EventGraph**

Although difficult to read, figure 7 gives an overview of a small part of the nodes layout in AR3_Regie blueprint required to run the child version of it, shown in figure 2 and used by a non computer-savvy user. Each blueprint is a C++ class that must be placed in the level, as an "instance", to execute its instructions. There are two types of blueprints, depending on how they are placed in a level. Either it is placed directly in the level, in which case it is called an Actor, or it is added to an existing Actor, in which case it is called an Actor Component or simply Component. GC, GA and GP objects are respectively CineCameraActor, SkeletalMeshActor and TargetPointActor blueprint classes, in the UE terminology, that are "instanciated" in the level, which basically means that they are drag&drop in the level interface.

The AKN_Regie user "programs" in the child blueprint Regie_Cuesheet, which we will call AR3_Regie_Child to focus on its kinship with AR3_Regie parent blueprint, whose functionality and architecture we'll now give an overview of. AR3 means that this is the third version of the **A**KN_**R**egie tool.

### 4.3 EventDispatchers in AR3_Regie blueprint

AR3_Regie's main functionalities are as follows:
- management of external peripherals,
- spawning of avatars and props,
- implementation of a node architecture for dispatching events,
- launch and execution of cues containing effects to be applied to spawned elements.

External peripherals are managed using Actor or Component blueprints containing the specific programming for each device. To retrieve MIDI signals from peripherals connected to the computer, one needs to activate the MIDI Device Support plugin, which is part of the Built-In plugins collection present by default in UE. Current supported MIDI device is the NanoKontrol controller manufactured by Korg. The gamepads used with AKN_Regie are Microsoft Xbox controllers directly recognized by UE using a different protocol. Up to 4 gamepads can be recognized simultaneously on a single computer.

The spawned Components and Actors send all the events emitted by the MIDI controllers, the gamepads (Actor AKN_GamePd_Input) and the keyboard (Actor AKN_Keyboard_Input) to three main EventDispatchers named AKN_NanoK_Regie, AKN_Gamepad_Regie and AKN_Keyboard_Regie, which transmit them throughout the AKN_Regie architecture. An EventDispatcher is an UE functionality allowing information transfer through a game.

AR3_Regie is then responsible for generating avatars and props for the chosen cast, and interprets the programmed devices configuration to perform two main tasks:

- activate go or goback cue in the cuelist, starting from the initialization cue; this is controlled by specifically programmed keys on a NanoKontrol or the keyboard;
- transmit translation or rotation using the keyboard or gamepad buttons on an avatar or prop.

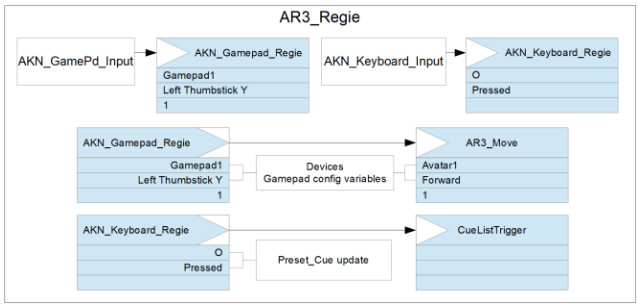

**Figure 8: EventDispatcher AR3_Regie architecture**

Figure 8 gives an insight of how the information circulates in the blueprint with EventDispatchers. [14] details the processes. The elements correspond to AKN_Regie's original functionalities for controlling avatars driven in real time by motion sensors in the AvatarStaging setup.

## 4.5 Calculating offset in blueprint

To illustrate the difference between organizing AR3_Regie_Child by a non-programmer user and programming AR3_Regie parent blueprint, we explain the offset functionality in AKN_Regie, central for directing avatars, even if very common in geometrical issues. The offset calculation for placing an avatar on a goal required specific development linked to the nature of the motion capture data, which are tweaked in real time. For the user, it simply means to use SetAvatar in a cue and affect the appropriate GA in the goal avatar parameter. Figure 9 proposes a hand-written sketch to geometrically move an avatar in a level. The artistic requirements of this repositioning and the impact on the staging are documented in [18]. The mathematical skills in geometry to calculate the offset vector to apply are of an elementary level.

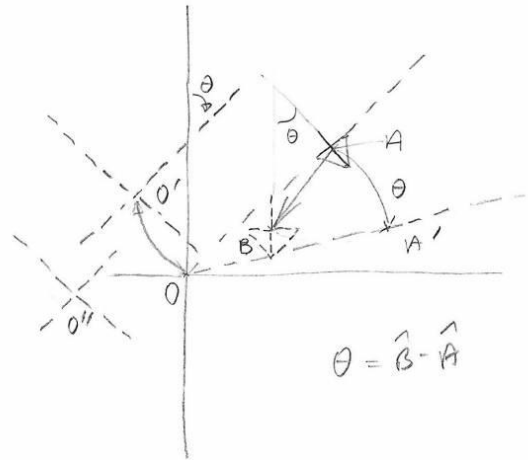

**Figure 9: Geometric transformation of avatar reference frame**

When retrieving motion capture data from an avatar located at A, the positional and rotational information of the avatar's skeletal joints is calculated relative to an origin O, which remains fixed. Repositioning the avatar indicated by an arrow at A to a new position indicated by the dotted arrow at B consists in practice of applying a rotation of angle θ to the point A that comes at A', then a translation $\overrightarrow{A'B}$. This is equivalent to making a reference frame change from O to O'' with a rotation of angle θ.

The programming of the operation requires a minimum Blueprint language skills as shown in figure 10. The implementation of the TransformOffset function in the SetAvatar node allow to precisely reposition any mocaptor controlled avatar in a level. This calculation is also used to rotate the avatar on itself.

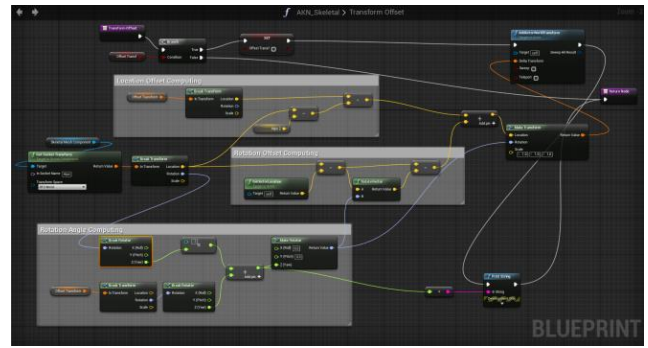

**Figure 10: Blueprint transcription of offset calculation by reference frame change**

## 4.6 Advanced features

The second version of AKN_Regie integrates the LiveLink functionality introduced in UE version 4.18. This feature facilitates the relationship with third-party motion capture software such as Motive (for the Optitrack optical system) or Axis Studio (for the Perception Neuron inertial systems), as well as with Motion Builder (developed by Autodesk) for motion retargeting issues. These developments have led to the creation of ready-to-use avatar libraries with AKN_Regie, and a method for equipping any avatar importable into UE and making it controllable with AKN_Regie.

The third version of AKN_Regie incorporated the development of a specific animation player, the Salient-Idle Player, which enables fine control of avatars without using real-time motion capture, but by arranging pre-recorded animations with an approach combining theater and video games [16].

It is worth noting that developing these functionalities in AR3_Regie parent blueprint requires a confirmed level of programming with Blueprint visual language.

## 5 Evolution of the tool along with productions

Currently on its third stage of development [19], AKN_Regie has been used in 10 performances staged by 7 directors as well as in multiple workshops in France, The United Kingdom, Greece, and Ukraine.

### 5.1 AKN_Regie foundation

The first stage took place between 2016 and 2018 and consisted in the foundation and the adaptation of the system. After a long period of collaboration between 2013 and 2016, Georges Gagneré, theatrical director and Cédric Plessiet, digital artist, came up with a first version of the plugin. It contained the main features that were meant to approach the avatars' direction to the physical actors' direction and simplify the comprehension of the digital universe to the theatre makers. In particular, a digital operator was able to position an avatar in the virtual space, turn the light on and off gradually or in one click (as we are used to do in the classical theatre), but as well could place a virtual camera and have an additional control of the camera and the avatar with a joystick.

This version was tested in practice in the framework of project *Masks and Avatars* [15] [Pluta]. Between 2016 and 2018 in Paris, Warwick, and Athens, 3 directors created 4 performances based on ancient Greek texts or improvisations with 2 live-acting mocaptors. The directors were challenged to rely the technical and digital part of their performances on the people who had no previous experience in 3D art and programming. During the experience the plugin was reconfigured to become more intuitive and comfortable in use on the one hand, and on the other to embrace multiple wished new features, like a possibility of rapid changing of virtual decorations, various use of props and VFX.

In summer of 2018 a second version of the plugin was released with a detailed documentation of the programming architecture on one side, and user step-by-step tutorials on the other side. Available on [16] the documentation is addressed to the theatrical students and theatre-makers who want to integrate the system in their work and for those who are willing to improve it within Blueprint environment.

The documentation and the second version were successfully tested between the 2018 and 2019 within the staging of *Patrick and Venus*, written and directed by Anastasiia Ternova where she invited two theatre-makers without programming skills to take a total responsibility for the digital part of the performance and they achieved a satisfying autonomy using the motion capture device and AKN_Regie [39] (figure 11).

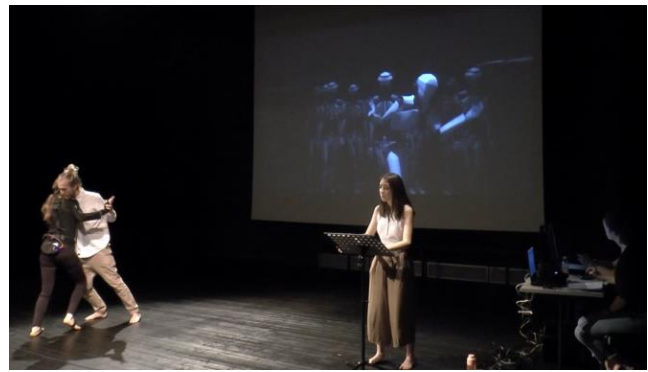

**Figure 11:** *Patrick and Venus*, by Ternova, 2019

In the beginning of 2019, the LiveLink functionaly was integrated in AKN_Regie and gave a possibility to use any kind of motion capture system (compatible with LiveLink) with any kind of biped avatars, while the previous version was limited to only one motion capture device and the motion retargeting worked correctly only for one type of avatars. And in summer 2019, the integration of the Salient-Idle Player led to the third version for the creation of *The Shadow*.

### 5.2 AKN_Regie evolution

*The Shadow* is a fairy tale by Hans Christian Andersen directed by Gagneré [13] (figure 12). The performance is played by one actor and 5 avatars animated with pre-recorded movements controlled by the Salient-Idle player. After having performed a salient movement, each avatar is programmed to pass to a smooth idle action that it maintains for the necessary time until the next cue is triggered [16].

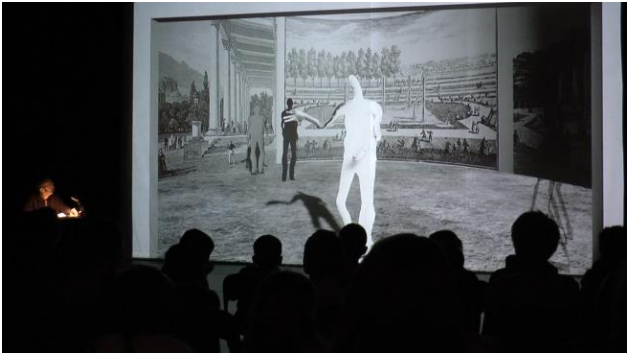

**Figure 12: *The Shadow*, after Andersen, directed by Gagneré, 2019**

This player is also used for the creation of *The Wizard without Shadow*, a Scottish fairy tale, adapted and directed by Anastasiia Ternova in 2021 [38] (figure 13) and *Les Solitudes de Donald Crowhurst*, written and directed by Cécile Roque-Alsina [39].

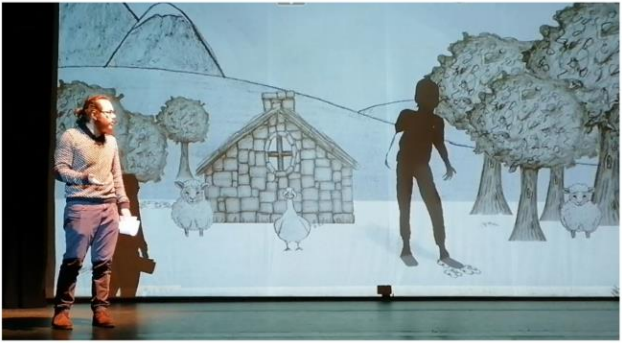

**Figure 13: *The Wizard without Shadow*, by Ternova, 2021**

Each project helped to supplement AKN_Regie plugin with new features that enriched its possibilities. For *The Shadow* it was needed to create a bond with Max/MSP musical software through Open Sound Control protocole, as the actor performed with different voice effects, and it was essential that he triggered cues at the same time as the voice modulations. As well the director wanted cinematic camera effects and specific object manipulations, so there was added a SetSequence node permitting to play frames of a special sequence that contained those pre-programmed effects.

*The Wizard without Shadow* is a performance for children starting from 5 years old and followed by interactive dialogues and games. To make it bright and exciting for kids, the staging team asked for many new features, as to restart particle system at any moment for magical effects, to add audio-files in the props and to create a live motion capture game, where a child playing in front of a Kinect caught or avoided virtual props. As this performance was created during COVID19 pandemic, it was also supposed that it could be played remotely. So, the team developed a special module to integrate a video of a life-acting actor in the remote virtual performance [38].

Finally for *Les Solitudes de Downald Crowhurst*, the director asked for a possibility to place the camera on the avatar's body. This simple action made the spectator feel being in a virtual reality headset and discover a strange reality of a mad man, who sails all alone in the middle of nowhere together with the avatar.

The result is that the plugin remains very flexible and integrates more and more creative tools that constantly increase the artistic capacities of the directing of avatars and virtual staging.

## 6 Discussion

AKN_Regie has been used on about ten creations, most of them by artists manipulating it solely from the Cuesheet, with no knowledge of Blueprint programming. Each new project has generally led to the introduction of new functionalities in successive versions of AKN_Regie. The evolution of UE has also offered new expressive possibilities, which have led to further development of the tool. During these evolutions, it became apparent that there was a kind of divide in the use of the tool according to the degree of appropriation of the code.

### 6.1 Plugin and Blueprint perspectives

Figure 14 represents what we call a Plugin perspective on the tool, positioning it as a simplified programming window within the overall UE environment, whose underlying complexity is masked. A non-programmer user only needs to know how to install the game engine, then activate the plugin and follow the instructions for the avatar and props control functionalities (see section 3).

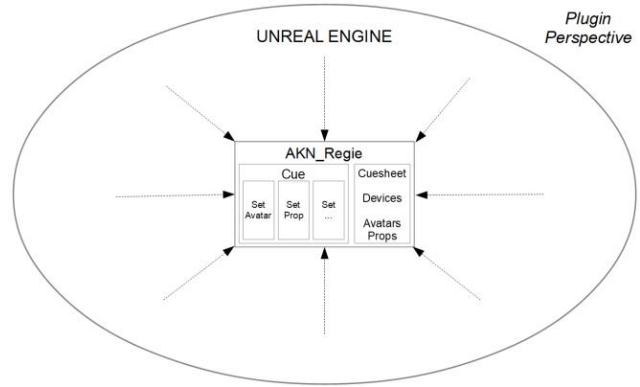

**Figure 14: Plugin perspective on the use of AKN_Regie**

There is also a blueprint programming documentation for AKN_Regie that explains how the user interface nodes are programmed using the engine's blueprint language and blueprints from other plugins added according to programming needs. As explained in section 4.1, these plugins are not required for standard engine operation. They are complementary programming elements providing access to new functionalities, such as the reception of information from third-party motion capture software (LiveLink plugin), or the ability to receive MIDI signals (MIDI Device Support plugin). Figure 15 positions AKN_Regie as a set of blueprints calling on these complementary blueprints and on other engine blueprints such as those produced with the Editor Animation interface for movements or the Editor Niagara interface for visual effects. AKN_Regie's blueprint architecture is thus explicit and potentially modifiable using the Blueprint programming language. It is no longer a black box appearing to a user in the form of a configurable plugin (figure2).

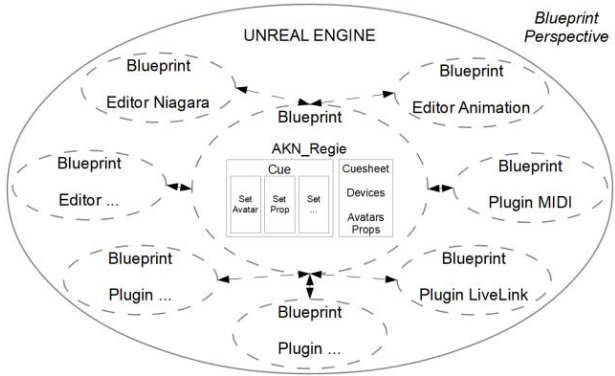

**Figure 15: Blueprint perspective on the use of AKN_Regie**

The Plugin perspective thus puts the solution of [37] into practice, encouraging artists to directly implement the technological dimension of their creative practice. Conversely, the Blueprint perspective enables the programmer to propose his programming ideas as artistic inputs. The juxtaposition of the two perspectives ultimately enables a team to share a common object around the direction of avatars on a mixed stage, as put forward by [22]. The non-programmer's recurring suggestions for creative improvements to the tool over the course of productions show that a dialogue exists between the two perspectives, encouraging the non-programmer to familiarize himself with programming and the use of complex tools such as video game engines, as recommended by [37]. Conversely, the programmer must integrate requests that originate outside the programming field. Finally, the circulation between the two perspectives favors the emergence or consolidation of the "proxy" function, as proposed by [3] and [33] as a guarantee of success for an interdisciplinary art, technology and science project.

We emphasize the importance of an environment such as the Blueprint visual language developed by Epic Games in UE. It makes the Blueprint perspective accessible and motivates us to strengthen programming documentation to facilitate the transition from Plugin to Blueprint. It is a solution to the problems raised by [28] and [1] that we have been able to implement with AKN_Regie.

## 6.2 C++ perspective

At this stage, we formulate the hypothesis that consists of taking programming complexity further and proposes that the circulation between the Plugin perspective's theatrical driven approach and Blueprint perspective architecture be extended to the engine's own language. This would mean tackling AKN_Regie's blueprints, those of complementary plugins such as LiveLink or MIDI, those produced with the Editor Anima or Niagara interfaces and all the engine's other functionalities at the C++ low programming level. As well as enabling an understanding of the primary nature of digital materials used creatively with human beings, the C++ perspective (figure 16) could facilitate, for example, a programming approach based on Design Patterns [20] and the use of the Unified Modeling Language UML [21]. It would optimize visual programming in blueprints and enable the integration of external C++ programming bricks into the existing UE code. Testing this hypothesis will be the next stage of our research.

The circulation of points of view between AKN_Regie's three perspectives requires each time to recontextualize knowledge in a different environment. Which approach is best suited to stimulating creative practice? We can assume *a priori* that the Plugin perspective facilitates a dialogue with the director and actors, but to the detriment of understanding the avatars' operating logic. We can also assume that the C++ perspective allows to invent new functionalities without guaranteeing their practical use on a theatrical stage with physical actors. The question boils down to finding a way to make cohabit these three different views on the use of the same digital materials for artistic creation.

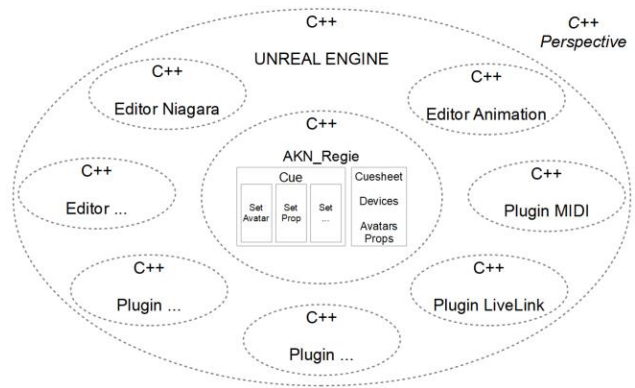

**Figure 16: C++ perspective on the use of AKN_Regie**

## 7 Conclusion

AvatarStaging framework consists in directing avatars on a mixed theatrical stage, enabling a co-presence between the materiality of the physical actor and the virtuality of avatars controlled in real time by mocaptors or specific animation players such as the Salient-Idle Player. It led to the implementation of the AKN_Regie software environment, programmed with the Blueprint visual language as a plugin for the UE video game engine. Its architecture is based on the AR3_Regie parent blueprint, whose main functionalities have been described, and enables the programming of the Regie_Cuesheet child blueprint, which constitutes the AKN_Regie user interface for non-programming artists.

The use of this interface requires a minimum of concepts for theatrical creations. The challenge is to program cues and parameterize external peripherals such as keyboards, MIDI controllers or gamepads, as well as a cast of avatars and props. It also offers advanced avatar direction features such as the Salient-Idle Player. The use of AKN_Regie by non-programming artists leads to new creative needs and new versions of AKN_Regie. We have described a blueprint programmer's view of the tool as Blueprint Perspective, and a non-programmer artist's as Plugin Perspective. We demonstrated that allowing a non-programmer to familiarize himself with the Blueprint perspective enrich the creative horizon and the quality of co-presence between avatars and actors.

We propose to extend the appropriation of AKN_Regie to the C++ perspective, to exploit the full potential of UE, coded in the same language. In parallel with the deployment of this C++ perspective, the next stage in AKN_Regie's development consists of taking facial motion capture into account and integrating UE's artificial intelligence tools such as finite state machines and behavioral trees to endow avatars with elements of scenic autonomy.

In Couchot's words, "the important thing is to live with these artificial beings, to make society with them. I believe that art can play a decisive role in this. Art can teach us to understand this exotic species, to take care of it, to tame it without dominating it, but also to derive pleasure from its company." [4] In the field of relations between actor, avatar and spectator, there is still much to be explored and realized to appropriate the code and weave new materialities and organicities that will help us progress in our humanity.

## ACKNOWLEDGMENTS

We would like to thank didascalie.net which supports the development of AvatarStaging framework (http://didascalie.net).

## REFERENCES

[1] Narges Ashtari, Andrea Bunt, Joanna McGrenere, Michael Nebeling, and Parmit K. Chilana. 2020. Creating Augmented and Virtual Reality Applications: Current Practices, Challenges, and Opportunities. In *Proceedings of the 2020 CHI Conference on Human Factors in Computing Systems* (Honolulu, HI, USA) (CHI '20). Association for Computing Machinery, New York, NY, USA, 1–13. https://doi.org/10.1145/3313831.3376722.

[2] Linda Candy and Ernest Edmonds. 2002. *Explorations in Art and Technology*. Springer, London, 8.

[3] Alexis Clay, Gaël Domenger, Julien Conan, Axel Domenger, and Nadine Couture. 2014. Integrating augmented reality to enhance expression, interaction & collaboration in live performances: A ballet dance case study. In *Proceedings of the 2014 IEEE International Symposium on Mixed and Augmented Reality-Media, Art, Social Science, Humanities and Design* (ISMAR-MASH'D). Munich, Germany, 10–12 September 2014, IEEE: Piscataway, NJ, USA, 21–29. https://hal.science/hal-00752227

[4] Edmond Couchot. 2018. Entretien de recherche. In *Interfaces numériques*, (7)1, 158-172.

[5] Edmond Couchot. 2022. *Automates, Robots et humains virtuels dans les arts vivants*. Presses Universitaires de Vincennes.

[6] Edmond Couchot and Norbert Hillaire. 2003. *L'art numérique. Comment la technologie vient au monde de l'art*. Paris, Edition Flammarion.

[7] Cycling74. Max/MSP. https://cycling74.com/products/max, accessed June 15, 2023.

[8] Andrea Davidson. 2016. Ontological shifts: Multi-sensoriality and embodiment in a third wave of digital interfaces. In *Journal of Dance & Somatic Practices*, 8(1), 21–42. https://doi.org/10.1386/jdsp.8.1.21_1

[9] Didascalie.net. AKN_Regie. https://avatarstaging.eu, accessed June 15, 2023.

[10] Epic Games. Blueprint. https://docs.unrealengine.com/5.2/en-US/blueprints-visual-scripting-in-unreal-engine/, accessed June 15, 2023.

[11] Epic Games. Unreal Engine. https://www.unrealengine.com/, accessed June 15, 2023.

[12] Andreas Fischer, Sara Grimm, Valentine Bernasconi, Angelika Garz, Pascal Buchs, and al. 2016. Nautilus: Real-Time Interaction Between Dancers and Augmented Reality with Pixel-Cloud Avatars. In *28ième conférence francophone sur l'Interaction Homme-Machine*, Oct 2016, Fribourg, Switzerland, 50–57. https://hal.science/hal-01386445

[13] Georges Gagneré. 2020. The Shadow. In *Proceedings of the 7th International Conference on Movement and Computing* (MOCO '20). Association for Computing Machinery, New York, NY, USA, Article 31, 1–2. https://doi.org/10.1145/3401956.3404250

[14] Georges Gagneré. 2022. AKN_REGIE, un plugin Unreal Engine pour la direction d'avatar sur une scène mixte. In *Actes des Journées d'Informatique Théâtrale*, octobre 2022, Lyon. https://inria.hal.science/hal-04152485

[15] Georges Gagneré, Andy Lavender, Cédric Plessiet, and Tim White. 2018. Challenges of movement quality using motion capture in theatre. In *Proceedings of the 5th International Conference on Movement and Computing* (MOCO '18). Association for Computing Machinery, New York, NY, USA, Article 44, 1–6. https://doi.org/10.1145/3212721.3212883

[16] Georges Gagneré, Tom Mays, and Anastasiia Ternova. 2020. How a Hyper-actor directs Avatars in Virtual Shadow Theater. In *Proceedings of the 7th International Conference on Movement and Computing* (MOCO '20). Association for Computing Machinery, New York, NY, USA, Article 15, 1–9. https://doi.org/10.1145/3401956.3404234

[17] Georges Gagneré and Cédric Plessiet. 2018. Experiencing avatar direction in low cost theatrical mixed reality setup. In *Proceedings of the 5th International Conference on Movement and Computing* (MOCO '18). Association for Computing Machinery, New York, NY, USA, Article 55, 1–6. https://doi.org/10.1145/3212721.3212892

[18] Georges Gagneré, Cédric Plessiet and Rémy Sohier. 2018. Interconnected Virtual Space and Theater: A Research–Creation Project on Theatrical Performance Space in the Network Era. In *Challenges of the Internet of Things* (eds I. Saleh, M. Ammi and S. Szoniecky). https://doi.org/10.1002/9781119549765.ch8

[19] Georges Gagneré and Cédric Plessiet. 2023. Quand le jeu vidéo est le catalyseur d'expérimentations théâtrales (2014-2019). In *Le jeu vidéo au carrefour de l'histoire, des arts et des médias*, dir. C. Devès. Lyon, Les Éditions du CRHI, 209-219.

[20] Erich Gamma, Richard Helm, Ralph Johnson, and John Vlissides. 1994. *Design Patterns. Elements of Reusable Object-Oriented Software*. Addison-Wesley.

[21] Ivar Jacobson, Grady Booch, and James Rumbaugh. 2000. *Le processus unifié de développement logiciel* (trad. de l'anglais par Zaim, V.). Paris : Eyrolles.

[22] Christian Jacquemin, Georges Gagneré, and Benoît Lahoz. 2011. Shedding light on shadow: real-time interactive artworks based on cast shadows or silhouettes. In *Proceedings of the 19th ACM international conference on Multimedia* (MM '11). Association for Computing Machinery, New York, NY, USA, 173–182. https://doi.org/10.1145/2072298.2072322

[23] Derrick de Kerckhove. 1980. A theory of Greek tragedy. In *SubStance*, Vol. 9, No. 4, Issue 29 (1980). The Johns Hopkins University Press, 23-36. https://doi.org/10.2307/3684038

[24] Myron W. Krueger. 1977. Responsive environments. *In Proceedings of the June 13-16, 1977, national computer conference* (AFIPS '77). Association for Computing Machinery, New York, NY, USA, 423–433. https://doi.org/10.1145/1499402.1499476
[25] Myron W. Krueger. 1998. I-Met-A-Morph. In *ACM SIGGRAPH 98 Electronic art and animation catalog* (SIGGRAPH '98). Association for Computing Machinery, New York, NY, USA, 81. https://doi.org/10.1145/281388.281651
[26] Norma Loewen. 1975. *Experiments in Art and Technology: A Descriptive History of the Organization*. Ph.D, diss., New York University.
[27] Leta E. Miller. 2001. Cage, Cunningham, and Collaborators: The Odyssey of Variations V. In *Musical Quarterly* 85, 3 (2001), 545–567. https://doi.org/10.1093/musqtl/85.3.545
[28] Michael Nebeling and Maximilian Speicher. 2018. The Trouble with Augmented Reality/Virtual Reality Authoring Tools. In *IEEE International Symposium on Mixed and Augmented Reality Adjunct* (ISMAR-Adjunct). Munich, Germany, 33-337. https://doi.org/10.1109/ISMAR-Adjunct.2018.00098
[29] Walter J. Ong. 2002. *Orality and Literacy – The Technologizing of the Word*. Routledge.
[30] Robin Oppenheimer. 2011. *The strange dance : 9 evenings : theatre & engineering as creative collaboration*. Ph.D, diss., Simon Fraser University. https://summit.sfu.ca/item/11626
[31] Izabella Pluta. 2019. When Theater Director Collaborates with Computer Engineer. In *Emerging Affinities - Possible Futures of Performative Arts*, dir. Borowski, M.; Chaberski, M.; Sugiera, M. Theater, V.127; Transcipt Verlag: Bielefeld, 49-70. https://doi.org/10.1515/9783839449066-004
[32] Franck Popper. 1975. *Art, Action and Participation*. Studio Vista and New York University Press.
[33] Cristina Portalés. 2018. When Augmented Reality Met Art: Lessons Learned from Researcher–Artist Interdisciplinary Work. In *Multimodal Technologies and Interaction*, 2(2), 17. https://doi.org/10.3390/mti2020017
[34] Processing. https://processing.org/, accessed June 15, 2023.
[35] Chris Salter. 2010. *Entangled*. MIT Press.
[36] Gilbert Simondon, 1958. *Du mode d'existence des objets techniques*. Éditions Aubier. [translated by the authors].
[37] Maja Stark, Elisabeth Thielen, Chrisoph Holtmann, André Selmanagić , Michael Droste and Leonid Barsht. 2022. XR Art and Culture: Successful Collaborations in Interdisciplinary Development Processes. In *i-com*, 21(1), 123-138. https://doi.org/10.1515/icom-2022-0011
[38] Anastasiia Ternova and Georges Gagneré. 2022. The Wizard Without Shadow: Staging and Evolution of a Performance Involving Motion Capture. In *International Journal of Art, Culture, Design, and Technology (IJACDT)*, vol.11, no.3, 1-13. https://doi.org/10.4018/IJACDT.314784
[39] Anastasiia Ternova and Georges Gagneré. 2023. Le potentiel créatif du plugin AKN_Regie dans un contexte théâtral. In *JIT 2022 - Journées d'Informatique Théâtrale*, Oct. 2022, Lyon. https://inria.hal.science/hal-04154382